\documentclass[conference]{IEEEtran}
\IEEEoverridecommandlockouts
\usepackage{cite}
\usepackage{amsmath,amssymb,amsfonts}
\newtheorem{definition}{Definition}
\usepackage{algorithmic}
\usepackage{graphicx}
\usepackage{subcaption}
\usepackage{textcomp}
\usepackage{xcolor}
\usepackage{seqsplit}
\usepackage{tikz}
\usepackage{tabularx}
\usepackage{array}
\usepackage{listings}
\usepackage{adjustbox}
\usepackage{multirow}
\usepackage{pgfplots}
\usepackage{comment}
\usepackage{makecell}
\usepackage{titlesec}
\titlespacing\section{0pt}{2pt plus 2pt minus 2pt}{0pt plus 2pt minus 2pt}
\titlespacing\subsection{0pt}{2pt plus 2pt minus 2pt}{0pt plus 2pt minus 2pt}
\titlespacing\subsubsection{0pt}{2pt plus 2pt minus 2pt}{0pt plus 2pt minus 2pt}

\pgfplotsset{width=10cm,compat=1.16}
\newcolumntype{M}[1]{>{\centering\arraybackslash}m{#1}}
\newcolumntype{P}[1]{>{\centering\arraybackslash}p{#1}}
\newcolumntype{C}[1]{>{\centering\let\newline\\\arraybackslash\hspace{0pt}}m{#1}}
\usetikzlibrary{shapes.geometric,shapes.symbols,fit,positioning,shadows, arrows.meta, backgrounds}
\renewcommand{\arraystretch}{1.0}
\def\BibTeX{{\rm B\kern-.05em{\sc i\kern-.025em b}\kern-.08em
		T\kern-.1667em\lower.7ex\hbox{E}\kern-.125emX}}
\begin{document}
	\makeatletter
	\newcommand{\linebreakand}{%
	\end{@IEEEauthorhalign}\vspace*{-0.3cm}
	\hfill\mbox{}\par
	\mbox{}\hfill\begin{@IEEEauthorhalign}
	}
	\makeatother
	\title{Joint Group Profiling and Recommendation via Deep Neural Network-based Multi-Task Learning
		%*\\ {\footnotesize \textsuperscript{*}Note: Sub-titles are not captured in Xplore and should not be used}
		\vspace{-0.4cm}
	}

	\author{
		
			\IEEEauthorblockN{Ngoc Luyen Le\textsuperscript{*\ddag}\\\textit{ngoc-luyen.le@hds.utc.fr}}
			\and
			\IEEEauthorblockN{Marie-Hélène Abel\textsuperscript{\ddag}\\\textit{marie-helene.abel@hds.utc.fr}}
			\linebreakand
			\IEEEauthorblockA{
				\textsuperscript{*} Gamaizer.ia, 93340 Le Raincy, France
			}
			\IEEEauthorblockA{
			\textsuperscript{\ddag} Université de technologie de Compiègne, CNRS, Heudiasyc (Heuristics and Diagnosis of Complex Systems),\\ CS 60319 - 60203 Compiègne Cedex, France
			}
			\vspace{-1cm}
	}

	%\author{\IEEEauthorblockN{1\textsuperscript{st} LE Ngoc Luyen}
		%\IEEEauthorblockA{\textit{Université de technologie de Compiègne, CNRS} \\
			%\textit{ Heudiasyc (Heuristics and Diagnosis of Complex Systems)}\\
			%CS 60319 - 60203 Compiègne Cedex, France \\
			%ngoc-luyen.le@hds.utc.fr}
		%\and
		%\IEEEauthorblockN{2\textsuperscript{nd} Marie-Hélène ABEL}
		%\IEEEauthorblockA{\textit{Université de technologie de Compiègne, CNRS} \\
			%\textit{ Heudiasyc (Heuristics and Diagnosis of Complex Systems)}\\
			%CS 60319 - 60203 Compiègne Cedex, France \\
			%marie-helene.abel@hds.utc.fr}
		%\and
		%\IEEEauthorblockN{3\textsuperscript{rd} Philippe GOUSPILLOU}
		%\IEEEauthorblockA{\textit{Vivocaz} \\
			%		%\textit{name of organization (of Aff.)}\\
			%8 B Rue de la Gare, 002200, Mercin-et-Vaux, France \\
			%p.gouspillou@vivocaz.fr}
		%}
	\maketitle

	\begin{abstract}
		
Group recommender systems aim to generate recommendations that align with the collective preferences of a group, introducing challenges that differ significantly from those in individual recommendation scenarios. This paper presents Joint Group Profiling and Recommendation via Deep Neural Network-based Multi-Task Learning, a framework that unifies group profiling and recommendation tasks within a single model. By jointly learning these tasks, the model develops a deeper understanding of group dynamics, leading to improved recommendation accuracy. The shared representations between the two tasks facilitate the discovery of latent features essential to both, resulting in richer and more informative group embeddings. To further enhance performance, an attention mechanism is integrated to dynamically evaluate the relevance of different group features and item attributes, ensuring the model prioritizes the most impactful information. Experiments and evaluations on real-world datasets demonstrate that our multi-task learning approach consistently outperforms baseline models in terms of accuracy, validating its effectiveness and robustness.

	\end{abstract}
	
	\begin{IEEEkeywords}
		Group Recommender System, Multi-task Learning, Group Profiling, Joint Learning, Representation Learning, Deep Neural Network
	\end{IEEEkeywords}
	
	\section{Introduction}

The explosive growth of online platforms and digital services has led to an overwhelming abundance of choices for users, making recommender systems an essential tool for filtering and personalizing content\cite{le2023personalized}. While traditional recommender systems primarily focus on individual users, many real-world scenarios involve group-based decision-making, such as choosing project for student groups, selecting movies for family viewing, or planning group travel activities \cite{masthoff2010group,quintarelli2016recommending,le2025context, le2023constraint}.

Group Recommender Systems (GRS) aim to provide recommendations that satisfy the preferences of a group of users rather than a single individual. This task introduces unique challenges not present in individual recommendation scenarios. One primary challenge is the aggregation of individual preferences into a coherent group preference that is acceptable to all or most group members \cite{baltrunas2010group}. Traditional approaches often rely on preference aggregation strategies like averaging individual preferences or selecting the least objectionable option, which may not adequately capture the complex dynamics of group decision-making \cite{berkovsky2010group}.

Another critical aspect of GRSs is group profiling, which involves creating a representation of the group's collective characteristics and preferences. Effective group profiling can enhance the recommendation process by considering the group's overall behavior patterns, social interactions, and the influence of individual members within the group \cite{huang2021novel}. However, existing methods typically treat group profiling and recommendation as separate tasks, potentially missing out on the benefits of shared information between them.

Recent advances in multi-task learning (MTL) have demonstrated significant improvements in various domains by leveraging shared information across related tasks \cite{ruder2017overview,dara2020survey}. However, the application of MTL to group recommender systems remains largely unexplored, particularly in the context of joint optimization of group profiling and recommendation tasks.

In this paper, we introduce \textit{Joint Group Profiling and Recommendation via Deep Neural Network-based Multi-Task Learning}, a MTL framework designed to simultaneously perform group profiling and recommendation within group recommender systems. MTL facilitates the joint training of these interrelated tasks, allowing the model to leverage shared representations, thereby improving generalization and performance. By treating group profiling as an auxiliary task to the primary recommendation task, our approach effectively captures the intricate relationships between group characteristics and item preferences. This unified framework not only enhances the understanding of group dynamics but also improves recommendation accuracy by uncovering latent features relevant to both tasks. To further refine the model, we integrate an attention mechanism that dynamically weighs the importance of various group features and item attributes, enabling the model to focus on the most impactful information for generating recommendations. Extensive experiments on real-world datasets confirm that our method outperforms baseline models, delivering superior accuracy and enhancing user satisfaction.

The remainder of this paper is organized as follows: Section 2 reviews related work in GRSs and MTL. Section 3 outlines the main contributions, detailing the approach and its components. Section 4 describes the experiments, including the datasets used for evaluation, and presents the results along with a discussion of the findings. Finally, Section 5 concludes the paper and suggests directions for future research.
	
	\section{Related Work}\label{section_relatedwork}
Our research intersects with several important areas in recommender systems and deep neural network-based learning. Therefore,  this section reviews relevant literature in group recommender systems, and multi-task learning approaches.
\subsection{Group Recommender Systems and Group Profiling}
Group recommender systems (GRS) have evolved significantly from early preference aggregation approaches to more sophisticated models that capture group dynamics. Early works like MovieLens \cite{miller2003movielens} or PolyLens \cite{o2001polylens} primarily focused on simple aggregation strategies such as averaging individual preferences or using minimum satisfaction guarantees. These approaches, while straightforward, often failed to capture complex group interactions and social dynamics.

More recent approaches have incorporated social relationship modeling into GRS. Guo et al. \cite{guo2015trustsvd} proposed SoGRec, a framework that integrates social connections and group preferences using a neural architecture to capture both individual and group-level social influences. Similarly, Cao et al. \cite{cao2019social} introduced AttentiveGroup, which uses attention mechanisms to model member interactions and their relative influence within groups.
Several studies have explored context-aware group recommendation. Quintarelli et al. \cite{quintarelli2016recommending} developed a system for ephemeral groups that adapts to different contextual situations. 

Group profiling has traditionally been approached through two main paradigms: aggregation-based and learning-based methods. Aggregation-based approaches, as surveyed by Masthoff \cite{masthoff2010group}, include strategies such as least misery, average satisfaction, and maximum pleasure. These methods, while interpretable, often oversimplify group dynamics.
Learning-based group profiling has gained prominence with the advent of deep learning. Sanchez et al. \cite{sanchez2012social} proposed a neural approach for modeling dynamic group preferences that adapts to changing group compositions and contexts. Hu et al. \cite{hu2013personalized} developed an attention mechanism to model group members' dynamic interactions and contributions. However, these approaches still treat profiling as a standalone task separate from recommendation.
\subsection{Multi-task Learning in Group Recommendation}
Recent advances in deep learning have introduced techniques for improving GRS, particularly through Multi-Task learning (MTL) and neural network architectures. MTL has shown promising results in various domains, as highlighted in surveys by Ruder \cite{ruder2017overview} and Crawshaw \cite{crawshaw2020multi}. In the context of recommender systems, MTL demonstrated improved performance by jointly learning user profiling and item recommendation \cite{li2020multi,wang2019multi}.

Several studies have applied advanced learning approaches to group recommendation scenarios. Chen et al. \cite{chen2021attentive} proposed a multi-task learning framework for group itinerary recommendation that jointly optimizes both personal and group preferences while capturing temporal dependencies. Their approach demonstrates how shared learning can improve the quality of group recommendations by leveraging the relationships between individual and group-level preferences.
Building on the power of neural networks, Feng et al. \cite{feng2022social} developed a deep neural network-based MTL approach that simultaneously learns social relationships and recommendation tasks, demonstrating the effectiveness of joint optimization in social-aware group recommendation. Their work shows how MTL can effectively capture complex social dynamics while maintaining computational efficiency.
Further advancing these approaches, Huang et al. \cite{huang2020efficient} introduced a neural network model that efficiently captures complex group dynamics through multiple specialized components, showing significant improvements in recommendation accuracy. Their work demonstrates how dedicated architectural designs can address the unique challenges of group recommendation.

These advanced approaches have made significant progress in addressing key challenges in group recommendation, including capturing dynamic group member interactions, balancing individual and group preferences, modeling contextual influences, and learning shared representations across different tasks. Despite these advances, most existing approaches treat group profiling and recommendation as separate tasks, potentially missing important interdependencies between them. Our work addresses this limitation by proposing a MTL framework that jointly optimizes both group profiling and recommendation tasks in the next section.
	\section{Joint Group Profiling and Recommendation: Our Approach}\label{proposition}

	In this section, we first formalize the problem of joint group profiling and recommendation using multi-task learning. Then, we present our deep neural architecture for carrying out these tasks simultaneously.
	\subsection{Task Formulation}
	Let define the fundamental components of our group recommender system:
	
	\begin{itemize} \item $\mathcal{G} = \{g_1, g_2, \ldots, g_N\}$: the set of groups. \item $\mathcal{U} = \{u_1, u_2, \ldots, u_M\}$: the set of users. \item $\mathcal{I} = \{i_1, i_2, \ldots, i_K\}$: the set of items. \item $\mathcal{R}$: the set of possible ratings or feedback scores. \item \( r_{g,i} \in \mathcal{R} \): Observed ratings provided by group \( g \) for item \( i \).
		 \item $\mathbf{x}_u \in \mathbb{R}^d$: feature vector representing user $u$.  derived from its members. \item $\mathbf{x}_i \in \mathbb{R}^f$: feature vector representing item $i$. \end{itemize}
%\item $\mathbf{x}_g \in \mathbb{R}^d$: feature vector representing group $g$,
	Each user \( u \in \mathcal{U} \) is represented by an individual feature vector \( \mathbf{x}_u \), which may include information such as preferences, interests, skills, and historical interactions. The objective is to leverage these individual user inputs to create meaningful group profiles and provide group-level recommendations.

	%Each group $g \in \mathcal{G}$ consists of a subset of users: $g \subseteq \mathcal{U}$. The group's features $\mathbf{x}_g$ are obtained by aggregating the features of its member users, typically through averaging or another suitable function. This aggregation captures the collective attributes and preferences of the group, serving as the foundation for both group profiling and recommendation tasks.
\begin{definition}[Group Profile]
	A group profile, denoted \( \mathbf{p}_g \in \mathbb{R}^h \), is a latent representation that encapsulates the collective preferences, characteristics, or behaviors of the group \( g \), derived from the aggregation of individual user inputs. This profile serves as a summary of the group's shared identity and is used to guide recommendations relevant to the group as a whole.
\end{definition}

Take the case of a university course where students are grouped into teams for a project-based assessment. Each student has a unique feature vector \( \mathbf{x}_u \) capturing their skills, interests, and academic history. Group profiles \( \mathbf{p}_g \) are created by aggregating these features, providing a snapshot of each group's overall capabilities and preferences.

\begin{definition}[Group Profiling Task] The group profiling task is the task of creating a representative summary, or latent profile, for a group by combining the characteristics, preferences, or behaviors of individual members. This profile captures the collective identity of the group, reflecting shared interests, capabilities, or preferences that guide decision-making for the group as a whole.
\end{definition}

 For instance, in a university course setting, group profiling for a team of students may involve averaging or aggregating each student's skills and interests to form a group profile. This group profile may then reflect a preference for projects focused on data science or software engineering, depending on the group's combined experience and interests.

\begin{definition}[Recommendation Task]
	The recommendation task involves predicting the rating \( \hat{r}_{g,i} \) that a group \( g \) would assign to an item \( i \). This prediction is based on both the group profile \( \mathbf{p}_g \) and the item's latent embedding \( \mathbf{e}_i \), and it serves to align recommendations with the group's collective interests.
\end{definition}

For example, for a student group profile that shows a preference for data science projects, the recommendation task might prioritize projects on machine learning, data visualization, or predictive analytics. The system ranks these projects based on predicted ratings or relevance scores, ultimately recommending the projects that best suit the group's profile.

Our approach addresses group recommendation through a MTL framework that jointly optimizes two key functions: group profiling and recommendation. The group profiling function learns to create group profiles by aggregating individual user characteristics and behaviors, generating a latent representation that captures the collective preferences of group members. The recommendation function then uses these learned group profiles along with item embeddings to predict how groups will rate different items. These two tasks are optimized simultaneously through a combined loss function that balances recommendation accuracy (measured by the difference between predicted and actual group ratings) and profiling quality. This joint optimization ensures that the learned group profiles are both meaningful representations of group characteristics and effective predictors of group preferences, while a hyperparameter controls the relative importance of each task during the learning process.

	Our goal is to recommend a list of top-$K$ items $\mathcal{I}g^{\text{top-}K}$ for each group $g$, based on the predicted ratings $\hat{r}_{g,i}$, while simultaneously learning meaningful group profiles $\mathbf{p}_g$. This dual objective ensures that the model not only predicts accurate recommendations but also gains a deeper understanding of the group's collective characteristics, which can be leveraged for various downstream applications.

\section{Deep Multi-Task Learning Architecture}

We introduce a Deep Neural Network-based Multi-Task Learning (DMTL) architecture which is designed to leverage individual user and item inputs for simultaneously learning both group profiles and generating group-level recommendations. This unified approach harnesses shared representations across tasks, enabling the model to capture nuanced user preferences within a group and deliver recommendations that resonate with the group's collective identity.

\subsection{Individual User and Item Embedding Layers}
Each user \( u \in \mathcal{U} \) and each item \( i \in \mathcal{I} \) are represented by their respective feature vectors \( \mathbf{x}_u \in \mathbb{R}^{d_u} \) and \( \mathbf{x}_i \in \mathbb{R}^{d_i} \), encompassing attributes such as preferences, skills, historical interactions, and item-specific characteristics. To transform these features into dense, meaningful representations, we employ separate embedding layers:
\begin{equation}
\mathbf{h}_u = \text{ReLU}\left(\mathbf{W}_{u} \mathbf{x}_u + \mathbf{b}_{u}\right)
\end{equation}
\begin{equation}
\mathbf{h}_i = \text{ReLU}\left(\mathbf{W}_{i} \mathbf{x}_i + \mathbf{b}_{i}\right)
\end{equation}
where  \( \mathbf{W}_{u}  \) and \( \mathbf{W}_{i} \) are the weight matrices, while \( \mathbf{b}_{u}\) and \( \mathbf{b}_{i} \) are the bias vectors for users and items, respectively. These transformations capture latent relationships within each feature space, setting the stage for effective interaction modeling.

\subsection{Attention Mechanism for Dynamic Weighting}
To create a meaningful interaction between user and item embeddings, we employ an Attention Mechanism Layer \cite{vaswani2017attention} that dynamically assigns importance weights to different aspects of the user-item pair. This mechanism operates as follows:
\begin{equation}
\mathbf{h}_{\text{concat}} = \mathbf{W}_{\text{concat}} \cdot \mathbf{h}_u + \mathbf{W}_{\text{concat}} \cdot \mathbf{h}_i + \mathbf{b}_{\text{concat}}
\end{equation}
\begin{equation}
\mathbf{h}_{\text{attn}} = \text{ReLU}\left(\mathbf{W}_{\text{attn}} \cdot \mathbf{h}_{\text{concat}} + \mathbf{b}_{\text{attn}}\right)
\end{equation}
\begin{equation}
\alpha_u = \text{Softmax}\left(\mathbf{W}_{\text{score}} \cdot \mathbf{h}_{\text{attn}} + \mathbf{b}_{\text{score}}\right)
\end{equation}
\begin{equation}
\mathbf{h}_{\text{attn\_item}} = \alpha_u \odot \mathbf{h}_i
\end{equation}
where \( \mathbf{W}_{\text{concat}}\) and \( \mathbf{b}_{\text{concat}} \) are the weight matrix and bias vector for concatenating user and item embeddings. \( \mathbf{W}_{\text{attn}} \) and \( \mathbf{b}_{\text{attn}} \) are the weight matrix and bias vector for the attention hidden layer. \( \mathbf{W}_{\text{score}}\) and \( \mathbf{b}_{\text{score}} \) are the weight matrix and bias vector for computing attention scores. \( \odot \) denotes element-wise multiplication.

The attention mechanism computes intermediate representations \( \mathbf{h}_{\text{attn}} \), which are then transformed into attention weights \( \alpha_u \) through a softmax function. These weights are applied element-wise to the item embeddings \( \mathbf{h}_i \), resulting in the attention-weighted item embeddings \( \mathbf{h}_{\text{attn\_item}} \).

\subsection{Combining Embeddings and Shared Representation Layer}
The attention-weighted item embeddings are combined with the original user embeddings as follows:
\begin{equation}
\mathbf{h}_{\text{combined}} = \mathbf{h}_u + \mathbf{h}_{\text{attn\_item}}
\end{equation}
This combined embedding \( \mathbf{h}_{\text{combined}} \in \mathbb{R}^{h_1} \) is then passed through a Shared Representation Layer, which further refines and enhances the group profile by capturing higher-order interactions:
\begin{equation}
\mathbf{z}_g = \text{ReLU}\left(\mathbf{W}_{\text{shared}} \cdot \mathbf{h}_{\text{combined}} + \mathbf{b}_{\text{shared}}\right)
\end{equation}

where \( \mathbf{W}_{\text{shared}}  \) and \( \mathbf{b}_{\text{shared}} \) are the weight matrix and bias vector for the shared dense layer.

The shared layer \( \mathbf{z}_g  \) serves as a robust foundation for both the classification (group profiling) and regression (rating prediction) tasks, ensuring that the learned representations are versatile and informative.

\subsection{Task-Specific Heads}

The architecture branches into two distinct Task-Specific Heads: (i) \textbf{Group Profiling Head}: This head generates the group profile \( \mathbf{p}_g^{\text{final}} \) by mapping the shared representation through a fully connected layer:
	\begin{equation}
	\mathbf{p}_g^{\text{final}} = \mathbf{W}_{\text{profile}} \cdot \mathbf{z}_g + \mathbf{b}_{\text{profile}}
\end{equation}
	where \( \mathbf{W}_{\text{profile}}  \) and \( \mathbf{b}_{\text{profile}}  \) are the weight matrix and bias vector for the profiling head. The resulting profile encapsulates the group's collective preferences and characteristics, providing a latent representation of the group's identity. (ii) \textbf{Recommendation Head}: This head predicts the rating \( \hat{r}_{g,i} \) for each item \( i \) based on the final group profile and the item's embedding:
	\begin{equation}
	\hat{r}_{g,i} = \mathbf{W}_{\text{rec}} \cdot \mathbf{p}_g^{\text{final}} + \mathbf{W}_{\text{item}} \cdot \mathbf{e}_i + \mathbf{b}_{\text{rec}}
\end{equation}
	where \( \mathbf{W}_{\text{rec}}  \) and \( \mathbf{b}_{\text{rec}}  \) are the weight matrix and bias vector for the recommendation head. \( \mathbf{W}_{\text{item}} \) is the weight matrix for the item embedding \( \mathbf{e}_i \).
	The linear combination of the group profile and item embedding produces a predicted rating \( \hat{r}_{g,i} \), indicating the group's anticipated interest in item \( i \).

\subsection{Multi-Task Loss Function}

To effectively train the model for both group profiling and recommendation, we employ a Multi-Task Loss Function that combines the losses from both tasks:
\begin{equation}
\mathcal{L} = \mathcal{L}_{\text{rec}} + \lambda \mathcal{L}_{\text{profile}}
\end{equation}
where  \( \mathcal{L}_{\text{rec}} \) is the recommendation loss, calculated as the Mean Squared Error (MSE) between actual and predicted ratings:
	\begin{equation}
	\mathcal{L}_{\text{rec}} = \frac{1}{|\mathcal{D}|} \sum_{(g,i,r) \in \mathcal{D}} (r_{g,i} - \hat{r}_{g,i})^2
\end{equation}
	where \( \mathcal{D} \) represents the set of observed group-item-rating tuples. \( \mathcal{L}_{\text{profile}} \) is the profiling loss, which evaluates the quality of the group profiles. This could be implemented using a classification loss such as cross-entropy loss if group profiling is treated as a classification task:
	\begin{equation}
	\mathcal{L}_{\text{profile}} = -\frac{1}{|\mathcal{G}|} \sum_{g \in \mathcal{G}} \sum_{c=1}^{C} y_{g,c} \log(\hat{y}_{g,c})
\end{equation}
	where \( y_{g,c} \) is the true label for group \( g \) in class \( c \), and \( \hat{y}_{g,c} \) is the predicted probability.
	\( \lambda \) is a hyperparameter that balances the contribution of the profiling loss relative to the recommendation loss, ensuring that neither task disproportionately influences the training process.

The proposed architecture effectively captures user characteristics, aggregates them into a group profile, and leverages this profile for group-level recommendations. By using user inputs, the model can adapt dynamically to the group's changing composition, enhancing both profiling accuracy and recommendation relevance.

	\section{Experiments}\label{experiments}
In this section, we introduce the dataset used to evaluate the performance of our proposed approach and describe the baseline models selected for comparison. We then present a detailed analysis of the experimental results.

\subsection{Dataset}

We evaluated our proposed approach on two distinct datasets (Table~\ref{tab00}): (i) ITM-Rec Dataset~\cite{zheng2023itm}: A domain-specific educational dataset collected from the Information Technology and Management department at Illinois Institute of Technology. The dataset encompasses student interactions and evaluations at both individual and group levels, providing a specialized context for academic recommendation tasks.
	(ii) MovieLens 100K\footnote{https://grouplens.org/datasets/movielens/100k/}: A benchmark dataset widely adopted in recommender systems research. While ITM-Rec focuses on educational resources, MovieLens represents entertainment preferences, enabling us to assess our group profiling methodology across different domains.

%\vspace{-0.1cm}
	\begin{table}[h]
		\renewcommand{\arraystretch}{1.2}
		\caption{Statistics on the datasets}\label{tab00}
		\begin{tabular}{|C{1.8cm}|C{0.8cm}|C{0.8cm}|C{0.8cm}|C{1.0cm}|C{1.0cm}|}
			\hline
			Dataset &  Nb of Users & Nb of Items & Rating Scale & Data Sparsity & Nb of Rating \\
			\hline
			ITM-Rec Dataset & 454 & 70  & [1,5] &  83.54\% & 5230\\ \hline
			Movielens 100K & 943 & 1682  & [1,5] &  93.70\% & 100,000\\ \hline
			
		\end{tabular}
		\vspace{-0.4cm}
	\end{table}
	
	\subsection{Baseline models and Evaluation Metrics}
	To evaluate our approach, we compare its performance with several established baseline methods, which serve as fundamental benchmarks for assessing algorithmic performance. These baselines, implemented within the Surprise Python library \cite{hug2020surprise} and the PyTorch framework \cite{paszke2019pytorch}, encompass a spectrum of approaches ranging from elementary to sophisticated: the baseline algorithm incorporates user and item biases; K-Nearest Neighbors variants (KNNBasic, KNNWithMeans)~\cite{koren2010factor}; matrix factorization methods including Singular Value Decomposition (SVD), SVD++~\cite{koren2008factorization}, Non-negative Matrix Factorization (NMF)~\cite{luo2014efficient}, and Slope One~\cite{lemire2005slope}. These methods, each supported by established research in collaborative filtering literature, collectively provide a framework for evaluating the performance gains achieved by our recommendation approaches.
	
	Regarding the evaluation metrics, we use precision, recall, and F1-score for the group profiling task. For the recommendation task, we employ two metrics: Precision@10 (P@10) and Recall@10 (R@10). P@10 measures the ratio of relevant items among the top-10 recommendations, formally defined as:
	\begin{equation}
		\text{P@}10 = \frac{|\text{relevant items} \cap \text{recommended items@}10|}{10}
	\end{equation}
	
	While R@10 quantifies the proportion of relevant items captured within the top-10 recommendations:
		\begin{equation}
\text{R@}10 = \frac{|\text{relevant items} \cap \text{recommended items@}10|}{|\text{total relevant items}|}
\end{equation}

	These metrics provide a assessment framework: P@10 evaluates recommendation accuracy, while R@10 measures the completeness of relevant item coverage. 
	
	\subsection{Experimental Results}
To evaluate the group profiling task, we applied KMeans clustering to create 20 groups for both datasets. The t-SNE visualizations (Figure \ref{fig_group}) illustrate the distribution of these groups, revealing distinct patterns. The Movielens 100K dataset exhibits well-separated clusters with minimal overlap, achieving strong performance metrics in precision, recall, and F1 score (Table \ref{tab01}), which indicates effective group profiling. In contrast, the ITM-Rec dataset shows more scattered and overlapping clusters in the t-SNE visualization but achieves even higher performance metrics, suggesting that despite the visual complexity, the dataset's underlying structure allows for highly accurate group predictions. This highlights the importance of leveraging high-dimensional relationships in group profiling tasks, which may not be fully apparent in 2D visualizations.

	\begin{figure}[htbp]
		\centering
		\begin{subfigure}[b]{\columnwidth}
			\centering
			\includegraphics[width=0.7\textwidth]{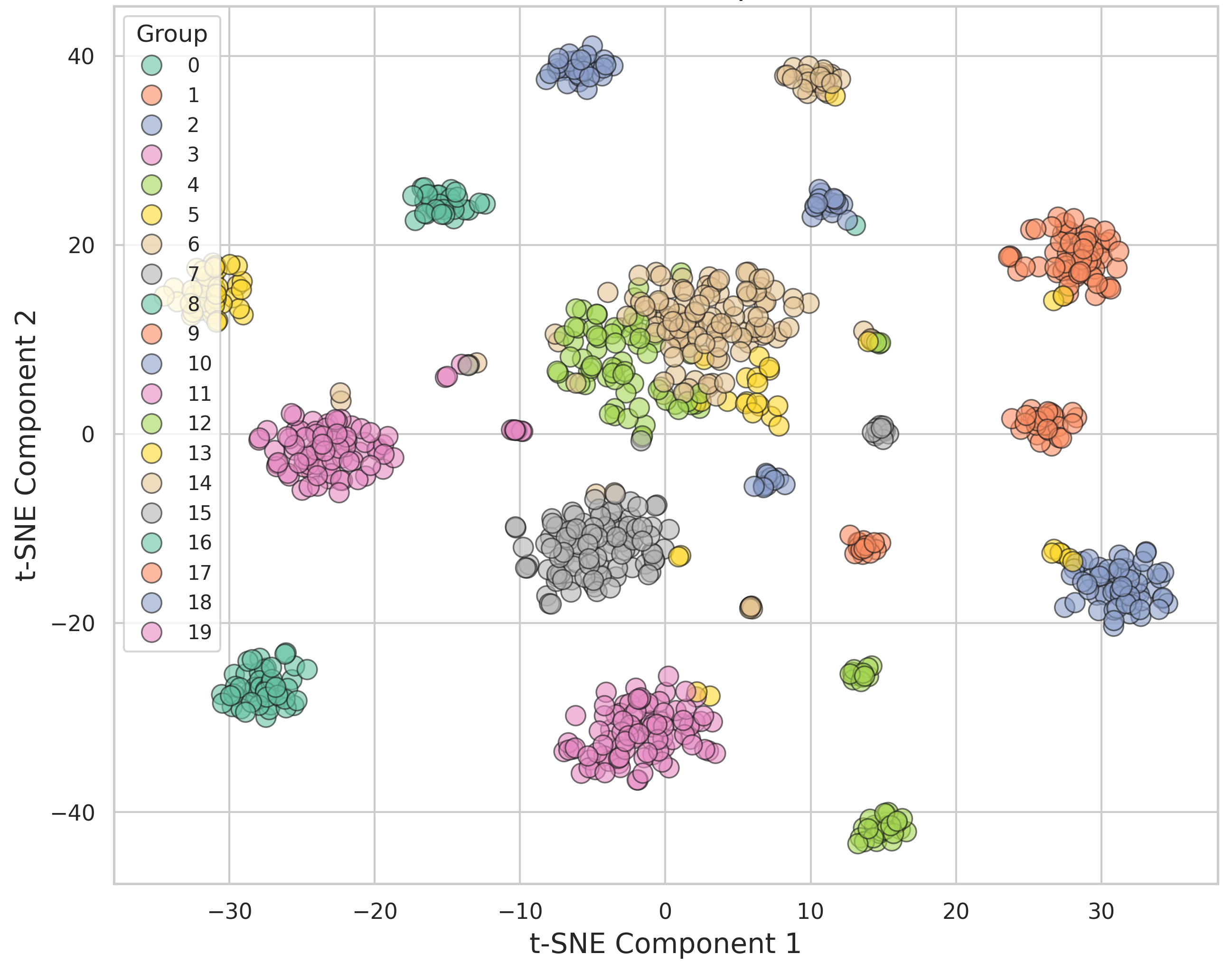}
			\caption{Groups visualization in the Movielen 100k}
			\label{fig_ml}
		\end{subfigure}
		\vspace{1cm}  % Vertical space between subfigures
		\begin{subfigure}[b]{\columnwidth}
			\centering
			\includegraphics[width=0.7\textwidth]{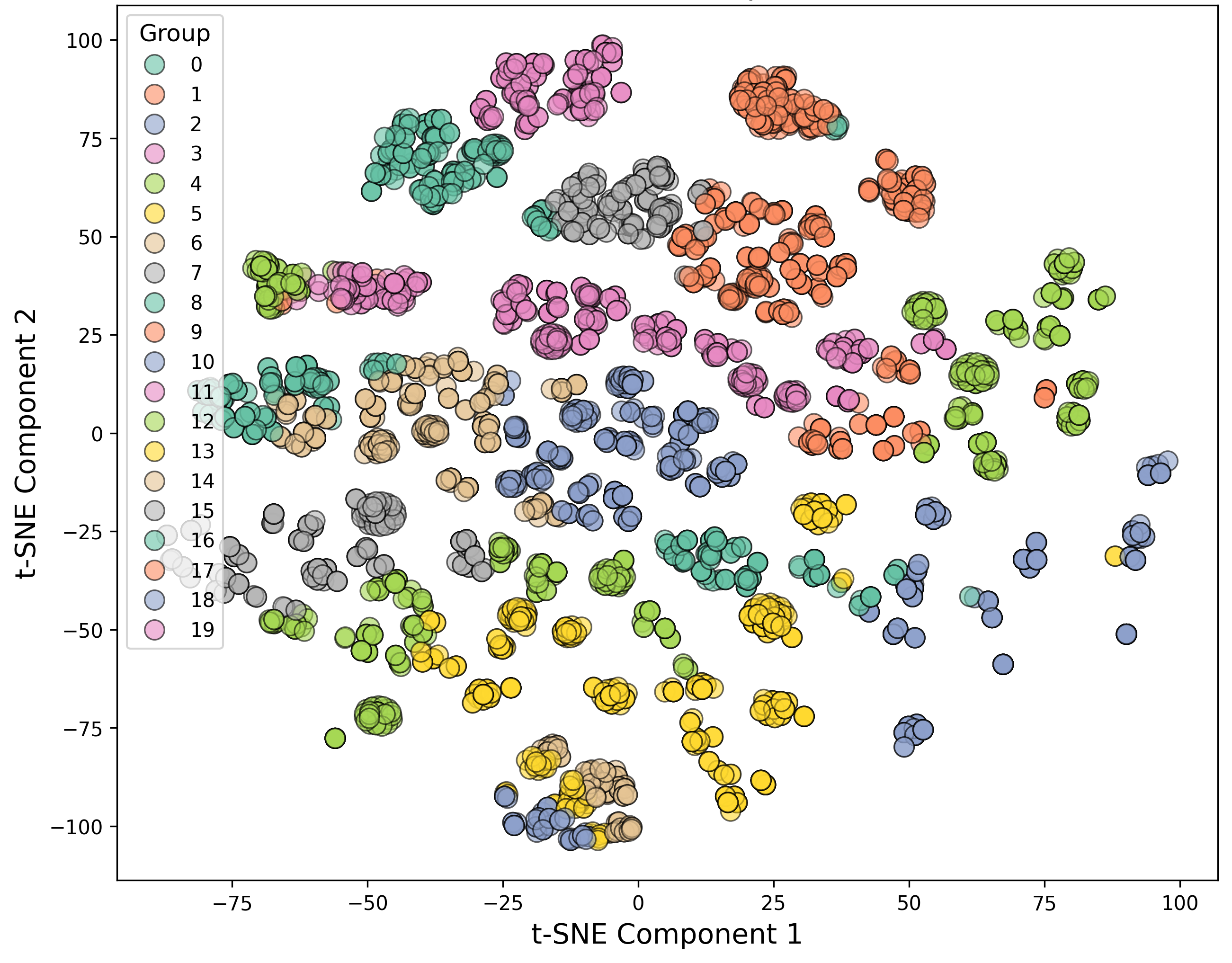}
			\caption{Groups visualization in the ITM-Rec Dataset}
			\label{fig_itm}
		\end{subfigure}
		\vspace{-1.4cm}
		\caption{t-SNE Visualization of different groups (User Clusters)}
		\label{fig_group}
	\end{figure}

	%\vspace{-0.4cm}
			\begin{table}[h]
		\renewcommand{\arraystretch}{1.2}
		\caption{Group profiling task results}\label{tab01}
		\centering
		\begin{tabular}{|C{2.0cm}|C{1.5cm}|C{1.5cm}|C{1.5cm}|}
			\hline
			Dataset &  Precision & Recall & F1 Score \\
			\hline
			Movielen 100K & 0.9059 & 0.9079  & 0.8893 \\ \hline
			ITM-Rec Dataset & 0.9754 & 0.9745  & 0.9740  \\ \hline
			
		\end{tabular}
	\end{table}
	
For the recommendation tasks across both datasets, our DMTL approach consistently outperforms all baseline and traditional methods, including KNN variants, SVD, SVD++, NMF, and Slope One as shown in the table \ref{tab02}.
Specifically, on the Movielens 100K dataset, DMTL achieves a significant performance boost with a P@10 of 0.6867 and a R@10 of 0.8063. This represents a marked improvement over the next best-performing method, SVD++, which records a P@10 of 0.5890 and an R@10 of 0.7248. These results highlight DMTL's superior ability to capture complex interactions within a densely populated dataset, enabling more accurate and relevant recommendations.

On the other hand, for the ITM-Rec dataset, despite encountering lower overall precision scores due to the limited sample size, DMTL still demonstrates a notable improvement. It achieves a P@10 of 0.3182 and an R@10 of 0.9306, outperforming SVD++ which has a P@10 of 0.3065 and an R@10 of 0.9241. The consistently higher recall indicates that DMTL is particularly effective at identifying a larger proportion of relevant recommendations, even in sparser or educational settings where data may be less abundant.

	\begin{table}[h]
		\renewcommand{\arraystretch}{1.5}
		\caption{Comparison of different recommendation approaches}\label{tab02}
		\begin{tabular}{|M{2.5cm}|M{1.0cm}|M{1.0cm}|M{1.0cm}|M{1.0cm}|}
			\hline
			\multirow{2}{*}{Approach} &  \multicolumn{2}{c|}{Movielen 100K} & \multicolumn{2}{c|}{ITM-Rec Dataset}\\\cline{2-5}
			& P@10 & R@10 & P@10 & R@10 \\\hline
			Baseline & 0.579043 &  0.719375 & 0.305357 &  0.922781 \\\hline
			
			KNNBasic (User-Based) & 0.584681 &  0.722181 & 0.304688 &   0.921933 \\\hline
			
			KNNWithMeans (User-Based) & 0.580745 &  0.719248 & 0.304018 &  0.921327\\\hline
			
			KNNBasic (Item-Based) & 0.559362 &   0.704650 & 0.298437 &  0.914993 \\\hline
			
			KNNWithMeans (Item-Based) &0.582128 &   0.720977 & 0.304018 &  0.921190 \\\hline
			
			SVD & 0.583723 &   0.721407 &0.304241 &  0.921511 \\\hline
			
			SVD++ & 0.589043 &   0.724757 & 0.306473 &  0.924077 \\\hline
			
			NMF & 0.578085 &   0.718161 &0.302232 &  0.918649
			 \\\hline
			
			Slope One & 0.578617 &  0.718489 & 0.303571 &  0.920048 \\\hline
			
			\textbf{DMTL (Ours)}  & \textbf{0.686744} &  \textbf{0.806285} & \textbf{0.318158}  & \textbf{0.930575} \\
			\hline
			
		\end{tabular}
	\end{table}
	%\vspace{-0.6cm}
	Overall, these results validate DMTL's capability to generalize across different domains, delivering robust performance enhancements in both precision and recall compared to traditional recommendation methods. The ability of DMTL to maintain high recall rates underscores its effectiveness in ensuring that relevant items are not missed, while its superior precision ensures that the recommendations remain highly relevant to users. This balance between precision and recall makes DMTL a powerful tool for diverse recommendation scenarios, from densely populated datasets like Movielens 100K to more specialized and sparse datasets like ITM-Rec.
	
	The effectiveness of DMTL approach in capturing complex interactions and contextual nuances across diverse dataset structures is clearly demonstrated by our experimental results. This enhanced performance is primarily due to DMTL's ability to leverage shared representations that simultaneously optimize both classification and regression tasks. Such inter-task synergy functions as an implicit regularizer, effectively mitigating overfitting and improving the model's generalization capabilities. Although the DMTL framework introduces increased model complexity and requires meticulous balancing of task-specific objectives, our empirical findings confirm its adaptability and robustness in delivering more precise recommendations. These outcomes not only underscore DMTL's potential to advance personalized recommendation systems but also pave the way for future enhancements in scalability, feature integration, and adaptive learning mechanisms, thereby fully exploiting its multifaceted advantages.

	\section{Conclusion and Perspectives}
	
	In this paper, we investigate the integration of Joint Group Profiling and Recommendation via Deep Neural Network-Based Multi-Task Learning. Our approach simultaneously learns group profiles and generates personalized recommendations by leveraging shared representations across both tasks. By incorporating both group-level and individual-level information, DMTL enhances the model's ability to capture intricate interactions and dependencies, resulting in improved performance in both group profiling and recommendation tasks. Experimental evaluations on real-world datasets, including Movielens 100K and ITM-Rec, demonstrate that our DMTL framework consistently outperforms established methods, underscoring its superior capability to deliver accurate and relevant recommendations while effectively navigating the complex interactions and contextual nuances inherent in group decision-making processes. The synergy achieved through DMTL not only facilitates the creation of group profiles but also acts as an implicit regularizer, mitigating overfitting and enhancing generalization. Consequently, the model maintains high recall rates and delivers precise recommendations even in sparser or more complex datasets. Looking ahead, future research will focus on expanding this approach to incorporate unstructured data sources, such as textual reviews and social media interactions, to further enrich group profiling and enhance recommendation accuracy. Additionally, addressing scalability challenges through optimized training algorithms and leveraging distributed computing will be essential for deploying DMTL-based GRSs in larger, real-time environments, thereby fully harnessing the multifaceted advantages of this innovative framework.

	%\begin{table}[htbp]
	%\caption{Table Type Styles}
	%\begin{center}
	%\begin{tabular}{|c|c|c|c|}
	%\hline
	%\textbf{Table}&\multicolumn{3}{|c|}{\textbf{Table Column Head}} \\
	%\cline{2-4} 
	%\textbf{Head} & \textbf{\textit{Table column subhead}}& \textbf{\textit{Subhead}}& \textbf{\textit{Subhead}} \\
	%\hline
	%copy& More table copy$^{\mathrm{a}}$& &  \\
	%\hline
	%\multicolumn{4}{l}{$^{\mathrm{a}}$Sample of a Table footnote.}
	%\end{tabular}
	%\label{tab1}
	%\end{center}
	%\end{table}
	
	%\begin{figure}[htbp]
	%\centerline{\includegraphics{fig1.png}}
	%\caption{Example of a figure caption.}
	%\label{fig}
	%\end{figure}

	%\bibliographystyle{plain}
	\bibliographystyle{IEEEtran}
	\bibliography{references} 
	
\end{document}